\documentclass[11pt,journal,draftcls,onecolumn,peerreviewca]{IEEEtran}
\usepackage{amsfonts}
\usepackage{amsmath}
\usepackage{amssymb}
\usepackage{mathrsfs}
\usepackage{stmaryrd}
\usepackage{bm}
\usepackage{graphicx}
\usepackage{color, soul}
\usepackage{stfloats}
\usepackage[amsmath,thmmarks]{ntheorem}
\usepackage{theorem}
\usepackage{subfigure}
\usepackage{ntheorem}
\usepackage{url}
\usepackage{hyperref}
\usepackage{float}
\usepackage{algorithm}
\usepackage{algorithmic}

\theoremheaderfont{\sc}\theorembodyfont{\upshape}
\theoremstyle{nonumberplain}
\theoremseparator{}
\theoremsymbol{\rule{1ex}{1ex}}

\usepackage{indentfirst}
\par
\usepackage{graphics}
\begin{document}
\title{One-Bit-Aided Modulo Sampling for DOA Estimation}
\author{Qi Zhang, Jiang Zhu, Fengzhong Qu, and De Wen Soh}
%\date{April 4, 2017}
\maketitle
\begin{abstract}
Modulo sampling has recently drawn a great deal of attention for cutting-edge applications, due to overcoming the barrier of information loss through sensor saturation and clipping. This is a significant problem, especially when the range of signal amplitudes is unknown or in the near-far case. To overcome this fundamental bottleneck, we propose a one-bit-aided (1bit-aided) modulo sampling scheme for direction-of-arrival (DOA) estimation. On the one hand, one-bit quantization involving a simple comparator offers the advantages of low-cost and low-complexity implementation. On the other hand, one-bit quantization provides an estimate of the normalized covariance matrix of the unquantized measurements via the arcsin law. The estimate of the normalized covariance matrix is used to implement blind integer-forcing (BIF) decoder to unwrap the modulo samples to construct the covariance matrix, and subspace methods can be used to perform the DOA estimation. Our approach named as 1bit-aided-BIF addresses the near-far problem well and overcomes the intrinsic low dynamic range of one-bit quantization. Numerical experiments validate the excellent performance of the proposed algorithm.
\end{abstract}


\section{Introduction}
\label{sec:intro}
Direction-of-arrival (DOA) estimation refers to the process of estimating the bearing of targets from the outputs of a receiving sensor array, which is an active field of research in signal processing \cite{tuncer2009classical}. Traditional algorithms for DOA estimation such as subspace-based methods, compressed sensing-based methods, atomic norm-based approaches, and Bayesian approaches require high-resolution samples from the sensor array hardware \cite{yang2018sparse, vaidyanathan2010sparse, liu2012efficient, mamandipoor2016newtonized}. However, in many real-world scenarios, acquiring such observations is often impractical or requires power-consuming hardware devices, and two typical situations are introduced in \cite{fernandez2022computational,feuillen2023unlimited}. One scenario involves the ambiguity of a signal's amplitude range, which could potentially exceed the dynamic range of the analog-to-digital converter (ADC) by a significant margin. Hence, measurements may be truncated, leading to information loss, and conventional algorithms may become ineffective under such circumstances. The other challenge is the near-far problem, which arises in scenarios involving two emitters, one considerably closer to the receiver than the other. Consequently, unless a high-bit ADC is utilized which is power-consuming, the sensor faces a dilemma: it can either prioritize the near-field emitter, resulting in the far-field emitter being submerged in quantization noise, or it can attempt to recover information from the far-field emitter, leading to the clipping of samples from the near-field emitter.

A strategy to address the sensor saturation issue in DOA estimation is based on the use of a 1-bit ADC, where measurements capture only the sign of the signal \cite{bar2002doa, liu2017one, sedighi2021performance}. However, this architecture faces limitations when dealing with the near-far problem, as the weak signal will be submerged in the quantization noise. Recently, the unlimited sampling framework (USF) has been proposed to address the above problem, where a modulo operator is applied before sampling, and the USAlg algorithm is proposed to recover the original signal form modulo samples with a recovery guarantee based on the oversampling assumption\cite{bhandari2017unlimited,bhandari2020unlimited}. In addition, USAlg has been extended to DOA estimation \cite{fernandez2022computational,fernandez2021doa}. For zero-mean random vectors, the integer-forcing (IF) decoder is proposed to obtain the true samples from modulo samples with a known covariance matrix and the performance is shown to approach information theoretical limits \cite{ordentlich2016integer, ordentlich2018modulo}, and the blind version where the covariance matrix is estimated by an empirical method is also proposed \cite{romanov2021blind}. Furthermore, the linear prediction method combined with the IF decoder with the performance guarantee given by the rate-distortion theory is proposed for jointly stationary processes with zero mean under modulo sampling with known autocorrelation functions \cite{ordentlich2018modulo}, and has been extended to the blind version \cite{weiss2022blind1, weiss2022blind2}. Moreover, algorithms based on the Fourier domain are also proposed for periodic signals by using Prony’s method \cite{bhandari2021unlimited} and projected gradient descent method \cite{azar2022residual, azar2022robust}. 
The hardware prototype enabling unlimited sampling is also developed, as initially presented in \cite{bhandari2021unlimited} and subsequently discussed in \cite{mulleti2023hardware}. To deal
with hysteresis and folding transients in practical hardware, a
thresholding approach with a recovery guarantee is proposed in \cite{florescu2022surprising}. 

\begin{figure*}[htb!]
		\centering
		\includegraphics[width=14cm]{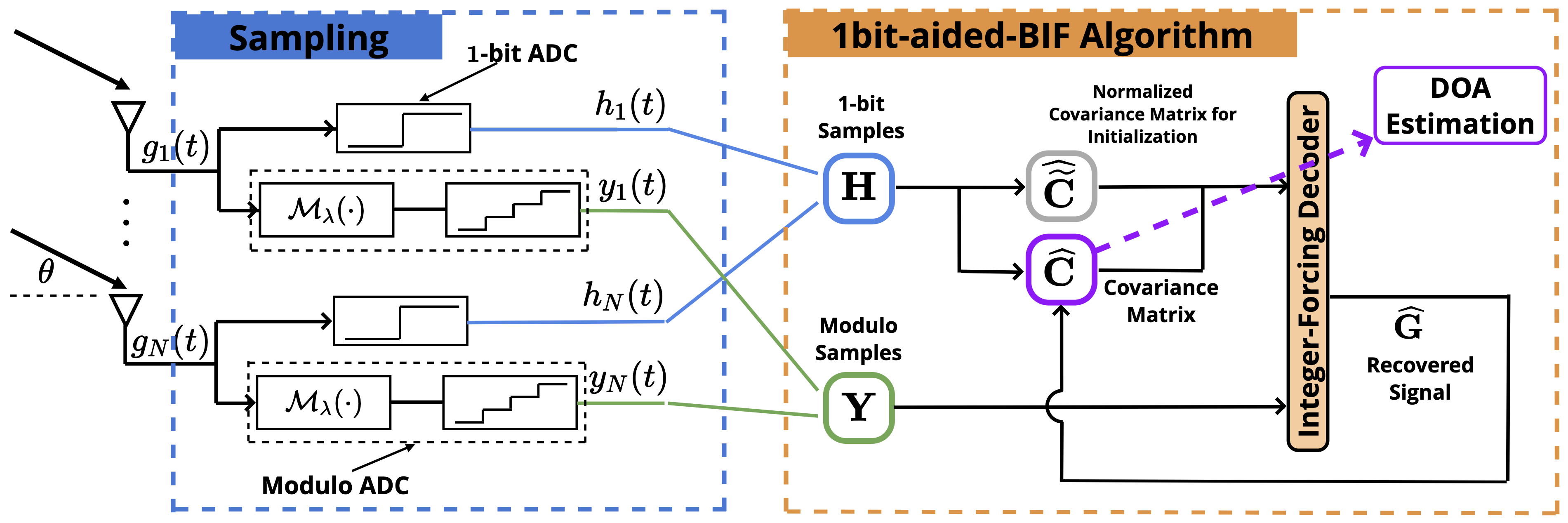}
		\caption{Computational array signal processing setup using the one-bit-aided modulo sampling/unlimited sensing architecture. Modulo non-linearity maps high-dynamic-range sensor array samples into low-dynamic-range folded samples. The one-bit quantization preserves the sign information of the original measurements. The 1bit-aided-BIF algorithm is used to carry out the estimation of DOAs based on the unified measurement scheme.}
        \label{1bitMSDOA}
\end{figure*}

Recently, there have been some works that combine modulo sampling with one-bit sampling. In \cite{graf2019one, eamaz2023unlimited, eamaz2022uno}, the one-bit modulo samples are obtained by applying one-bit sampling with time-varying thresholds after modulo sampling. In detail, to recover the original signal from one-bit modulo samples, the One-Bit US method is proposed by shrinking the dynamic range between the original signal and one-bit samples \cite{graf2019one} and the UNO technique is proposed by shrinking the dynamic range between the input signal and the time-varying sampling thresholds \cite{eamaz2023unlimited, eamaz2022uno}. In addition, an asynchronous implementation of one-bit sampling, i.e., the event-driven sampling architecture that incorporates a modulo nonlinearity prior to acquisition, together with a mathematically guaranteed recovery algorithm are proposed\cite{florescu2022time}. Unlike previous works, in this paper, we use both one-bit samples and modulo samples, which correspond to the most significant bit and the least significant $B$ bits of the original samples, to estimate the DOAs for uniform and sparse linear arrays, and the architecture is shown in Fig. \ref{1bitMSDOA}. Furthermore, with the aid of one-bit samples, we proposed an algorithm based on the IF decoder with unknown convariance matrix named as one-bit-aided blind IF (1bit-aided-BIF). In detail, we use one-bit samples to construct the normalized covariance matrix based on the arcsin law \cite{bar2002doa}. Then, the IF decoder is applied to recover the original signal. We will iteratively update the estimate of the covariance matrix using the recovered measurements that are consistent with the one-bit samples. Finally, subspace methods such as the root MUSIC algorithm are applied to perform the DOA estimation. Numerical results are conducted to demonstrate the performance of the proposed algorithm in the near-far problem.
\section{Problem Setup}
\label{sec:proset}
Consider linear arrays comprising $N$ sensors located at positions $\{\frac{d_n\lambda}{2}\}_{n=1}^N$, where $d_n\in\mathbb{N}$ is a non-negative integer, and $\lambda$ represents the signal wavelength. We define the set $\mathbb{D} = \{d_1,\cdots,d_N\}$ to describe the sensor spacing configuration.
For the $t$-th snapshot, the noisy unquantized measurement is given by
\begin{align}\label{measurements}
    \mathbf{g}(t) = \sum_{k=1}^K \mathbf{a}({\theta}_k){x}_{k}(t) + \mathbf{w}(t),~t= 1,\cdots,T,
\end{align}
where $\theta_k$ and $\mathbf{x}_{k}\triangleq [x_{k}(1),\cdots,x_{k}(T)]^{\rm T}\sim\mathcal{CN}({\mathbf 0},\sigma^2_k{\mathbf I}_T)$ are the DOA and complex amplitudes of the $k$-th target, $\mathbf{a}({\theta}_k)$ is the steering vector defined as
\begin{align}
    \mathbf{a}({\theta}_k) = \left[\mathrm{e}^{\mathrm{j} \pi d_1 \sin \theta_k}, \mathrm{e}^{\mathrm{j} \pi d_2 \sin \theta_k}, \ldots, \mathrm{e}^{\mathrm{j}  \pi d_N \sin \theta_k }\right]^{\mathrm{T}},
\end{align}
and $\mathbf{w}(t)\sim\mathcal{CN}(\mathbf{0},\sigma^2\mathbf{I})$ is the additive white Gaussian noise, and $\mathbf{w}(t_1)$ is independent of $\mathbf{w}(t_2)$ for $t_1\neq t_2$. In this paper, several prototypes of linear arrays are considered as follows. In the case of Uniform Linear Arrays (ULAs), the sensors are uniformly spaced, with $\mathbb{D} = \{0,1,\cdots,N-1\}$. For Coprime Arrays, the prototype set $\mathbb{D}$ is defined as $\{P q \mid 0 \leq q \leq P-1\} \cup\{Q p \mid 0 \leq p \leq P-1\}$, where $P$ and $Q$ are coprime integers, and $N = P+Q-1$. In the context of Nested Arrays, we have $\mathbb{D} = \{n\}_{n=1}^{N_1} \cup\left\{m\left(N_1+1\right)\right\}_{m=1}^{N_2}$, with $N = N_1 + N_2$.

Rather than acquiring the discrete-time signal $\mathbf{g}(t)$ generated by a high-dynamic-range and high-resolution ADC, we acquire the most significant bit from the $1$-bit ADC and the least significant $B$ bits from the modulo ADC in this paper. Specifically, the one-bit quantized sample of the sensor $n$ at time $t$ is given by
\begin{align}\label{1bitmeasure}
    h_{n}(t) = {\rm sign}(\Re\{{g}_{n}(t)\})/\sqrt{2} +{\rm sign}(\Im\{{g}_{n}(t)\})/\sqrt{2},
\end{align}
where $\Re\{{g}_{n}(t)\}$ and $\Im\{{g}_{n}(t)\}$ return the real part and imaginary part of ${g}_{n}(t)$ respectively. In addition, the $B$-bit quantized modulo sample from the $n$-th sensor at time $t$ is
\begin{align}
    y_{n}(t) = \mathcal{Q}_{B,\lambda}(\mathcal{M}_{\lambda}(\Re\{{g}_{n}(t)\})) + {\rm j}\mathcal{Q}_{B,\lambda}(\mathcal{M}_{\lambda}(\Im\{{g}_{n}(t)\})),
\end{align}
where
\begin{align}
    \mathcal{M}_{\lambda}(z) = z - 2\lambda\lfloor\frac{z}{2\lambda} + \frac{1}{2}\rfloor
\end{align}
is the modulo operator with range $[-\lambda,\lambda]$, and $\mathcal{Q}_{B,\lambda}(\cdot)$ is the $B$-bit quantization operator. Let $D = 2^B$, the quantization intervals of $\mathcal{Q}_{B,\lambda}(\cdot)$ are $
    \left\{\left(-\lambda+\frac{2\lambda l}{D}, -\lambda+\frac{2\lambda(l+1)}{D}\right)\right\}_{l=0}^{D-1},$
and for $z\in \left(-\lambda+\frac{2\lambda l}{D}, -\lambda+\frac{2\lambda(l+1)}{D}\right)$ in the $l$-th interval, the quantized value is $\mathcal{Q}_{B,\lambda}(z) = -\lambda + \frac{2\lambda(2l+1)}{D}$
which is the middle point of the interval. For the sake of notation convenience, we are using a uniform quantizer here. However, it is worth noting that the algorithm we propose later is equally applicable to non-uniform quantizers.

\section{Algorithm}
\label{sec:alg}
In this section, the 1bit-aided-BIF algorithm will be proposed to perform the DOA estimation given the one-bit observations and quantized modulo measurements.
\subsection{Estimate the Normalized Convariance Matrix}
\label{sec:estcov}
Let $\bar{\mathbf{C}}\triangleq \mathbb{E}[\mathbf{h}\mathbf{h}^{\rm H}]$ be the covariance matrix of one-bit samples (\ref{1bitmeasure}). Obviously $\bar{C}_{ii}=1$. According to the arcsine law, the correlation matrix $\widetilde{\mathbf{C}}$ (or the normalized covariance matrix) of unquantized measurements (\ref{measurements}) and the covariance matrix $\bar{\mathbf{C}}$ of one-bit measurements are related via \cite{bar2002doa}
\begin{align}\label{asinelaw}
     {\widetilde{\mathbf{C}}} = \sin\left(\frac{\pi}{2}\Re\left\{\bar{\mathbf{C}}\right\}\right)+{\rm j}\sin\left(\frac{\pi}{2}\Im\left\{\bar{\mathbf{C}}\right\}\right),
\end{align}
where $\sin(\cdot)$ is an element-wise operator. The empirical covariance matrix (also normalized covariance matrix) estimate of one-bit measurements is
\begin{align}\label{1bitcov}
\widehat{\bar{\mathbf{C}}}=\frac{1}{L}\sum_{l=1}^L\mathbf{h}_l\mathbf{h}^{\rm H}_l.
\end{align}
According to (\ref{asinelaw}) and (\ref{1bitcov}), an estimate of the normalized covariance matrix $\widehat{{\widetilde{\mathbf{C}}}}$ can be obtained.

As the IF decoder introduced later is specifically designed for real-valued signals, we will consider the real vector $\bar{\mathbf{g}}(t) = [\Re\{\mathbf{g}(t)\};\Im\{\mathbf{g}(t)\}]$ instead of $\mathbf{g}(t)$ below. In reality, based on the facts that $\mathbb{E}[\mathbf{g}\mathbf{g}^{\rm H}] = \mathbb{C}$ and $\mathbb{E}[\mathbf{g}\mathbf{g}^{\rm T}] = \mathbf{0}$, the covariance matrix ${\mathbf{C}}_r\triangleq \mathbb{E}[\bar{\mathbf{g}}\bar{\mathbf{g}}^{\rm T}]$ and ${\mathbf{C}}\triangleq \mathbb{E}[\mathbf{g}\mathbf{g}^{\rm H}]$ have the following relationship
\begin{align}\label{norconmat}
    {{\mathbf{C}}}_r =
        \left[\begin{array}{ll}
\Re\{{{\mathbf{C}}}\} & -\Im\{{{\mathbf{C}}}\} \\
\Im\{{{\mathbf{C}}}\} & \Re\{{{\mathbf{C}}}\}
\end{array}\right].
\end{align}
Thus the estimate of the nomalized convariance matrix of $\bar{\mathbf{g}}(t)$ denoted as $\widehat{\widetilde{\mathbf{C}}}_r$ can be obtained according to arcsine law and equation (\ref{norconmat}).
\subsection{Integer-Forcing Decoder}
Let $\bar{\mathbf{y}}(t) = [\Re\{\mathbf{y}(t)\};\Im\{\mathbf{y}(t)\} ]$, we have the fact that
\begin{align}
    \bar{\mathbf{y}}(t) = \bar{\mathbf{g}}(t) +  2\lambda\boldsymbol{\epsilon}(t) + \mathbf{z}(t),
\end{align}
where $\mathbf{z}(t) =  \bar{\mathbf{y}}(t) - \mathcal{M}_{\lambda}(\bar{\mathbf{g}}(t))$ is the quantization noise, and $\boldsymbol{\epsilon}(t) \in\mathbb{Z}^{2N}$ is an integer vector. Although $\mathbf{z}(t)$ is determined by $\bar{\mathbf{g}}(t)$, under suitable conditions, models that assume that each element of $\mathbf{z}(t)$ follows a uniform distribution $\mathcal{U}([-\frac{\lambda}{D},\frac{\lambda}{D}])$ provide good performance \cite{ordentlich2014precoded}. Note that for any integer matrix $\mathbf{A}\in\mathbb{Z}^{2N\times 2N}$, based on the fact that $\mathcal{M}_{\lambda}(2\lambda\mathbf{A}\boldsymbol{\epsilon}(t)) = \mathbf{0}$, we have
\begin{align}\label{modCom}
    \mathcal{M}_{\lambda}(\mathbf{A}(\bar{\mathbf{g}}(t) + \mathbf{z}(t))) = \mathcal{M}_{\lambda}(\mathbf{A}\bar{\mathbf{y}}(t)).
\end{align}
In the IF decoder, an invertible matrix $\widehat{\mathbf{A}}$ is obtained which aims to compress the amplitude of $\bar{\mathbf{g}}(t)$ by solving the optimization problem
\begin{align}\label{IFopt}
&\min _{\substack{\mathbf{A} \in \mathbb{Z}^{2N \times 2N} \\ \operatorname{det}(\mathbf{A}) \neq 0}} \max _{k=1, \ldots, 2N} \mathbb{E}\left(\left\|\mathbf{a}_k^{\rm T}(\bar{\mathbf{g}}(t) + \mathbf{z}(t))\right\|^2\right)\notag\\
    \iff &\min _{\substack{\mathbf{A} \in \mathbb{Z}^{2N \times 2N} \\ \operatorname{det}(\mathbf{A}) \neq 0}} \max _{k=1, \ldots, 2N} \mathbf{a}_k^T\left(\mathbf{C}_r+\frac{\lambda^2}{3D^2} \mathbf{I}_{2N}\right) \mathbf{a}_k,
\end{align}
where $\mathbf{a}_k$ is the $k$th column of $\mathbf{A}$. This optimization problem is NP-hard in general, and the Lenstra-Lenstra-Lovász lattice reduction (LLL) algorithm can be applied to approximately solve it in polynomial time \cite{lenstra1982factoring}. Given that $\mathcal{M}_{\lambda}(\widehat{\mathbf{A}}(\bar{\mathbf{g}}(t) + \mathbf{z}(t))) = \widehat{\mathbf{A}}(\bar{\mathbf{g}}(t) + \mathbf{z}(t))$ with high probability, eq. (\ref{modCom}) implies $\widehat{\mathbf{A}}(\bar{\mathbf{g}}(t) + \mathbf{z}(t)) = \mathcal{M}_{\lambda}(\widehat{\mathbf{A}}\bar{\mathbf{y}}(t))$ and $\bar{\boldsymbol{g}}(t)$ can be estimated as
\begin{align}\label{estmateg}
\widehat{\bar{\mathbf{g}}}(t) = \widehat{\mathbf{A}}^{-1}\mathcal{M}_{\lambda}(\widehat{\mathbf{A}}\bar{\mathbf{y}}(t)).
\end{align}
\subsection{1bit-Aided-BIF Algorithm}
As the covariance matrix $\mathbf{C}_r$ is not known in our setting, we use the LLL algorithm to solve an approximate optimization problem of (\ref{IFopt}) in the initialization step as shown below
\begin{align}\label{IFoptinit}
\widehat{\mathbf{A}} = \min _{\substack{\mathbf{A} \in \mathbb{Z}^{2N \times 2N} \\ \operatorname{det}(\mathbf{A}) \neq 0}} \max _{k=1, \ldots, 2N} \mathbf{a}_k^T\widehat{\widetilde{\mathbf{C}}}_r\mathbf{a}_k,
\end{align}
where $\widehat{\widetilde{\mathbf{C}}}_r$ is the estimate of the nomalized convariance matrix derived in Section \ref{sec:estcov}.
Then, we can obtain the estimates of $\bar{\mathbf{g}}(t)$ according to (\ref{estmateg}). Let $\mathbb{T} = \{t| \text{sign}(\widehat{\bar{\mathbf{g}}}(t))/\sqrt{2} = \bar{\mathbf{h}}(t)\}$ denote the set of time instances where the estimated samples match the sign of the ground truth, and the covariance matrix $\mathbf{C}_r$ can be estimated as
\begin{align}\label{estcon}
\widehat{\mathbf{C}}_r = \frac{1}{|\mathbb{T}|}\sum_{t\in{\mathbb{T}}}\widehat{\bar{\mathbf{g}}}(t)\widehat{\bar{\mathbf{g}}}(t)^{\rm T}.
\end{align}
Next, we update $\widehat{\mathbf{A}}$ by solving (\ref{IFopt}), where $\mathbf{C}_r$ is replaced by $\widehat{\mathbf{C}}_r$. We will iteratively update the estimate of the covariance matrix $\widehat{\mathbf{C}}_r$ and set $\mathbb{T}$ until $\widehat{\mathbf{C}}_r$ achieves convergence or the maximum number of iterations is reached. Finally, the covariance matrix of $\mathbf{g}(t)$ can be estimated as
\begin{align}\label{estcon2}
    \widehat{\mathbf{C}} = \frac{1}{|\mathbb{T}|}\sum_{t\in{\mathbb{T}}}\widehat{{\mathbf{g}}}(t)\widehat{{\mathbf{g}}}(t)^{\rm H}
\end{align}
where $\widehat{{\mathbf{g}}}(t)$ is the complex form of $\widehat{\bar{\mathbf{g}}}(t)$, and subspace-based algorithms such as root MUSIC can be applied to perform DOA estimation. In general, our algorithm is summarized in Algorithm \ref{1bitUSDOA_alg}.
\begin{figure*}[htb!]
		\centering
            \includegraphics[width=17cm]{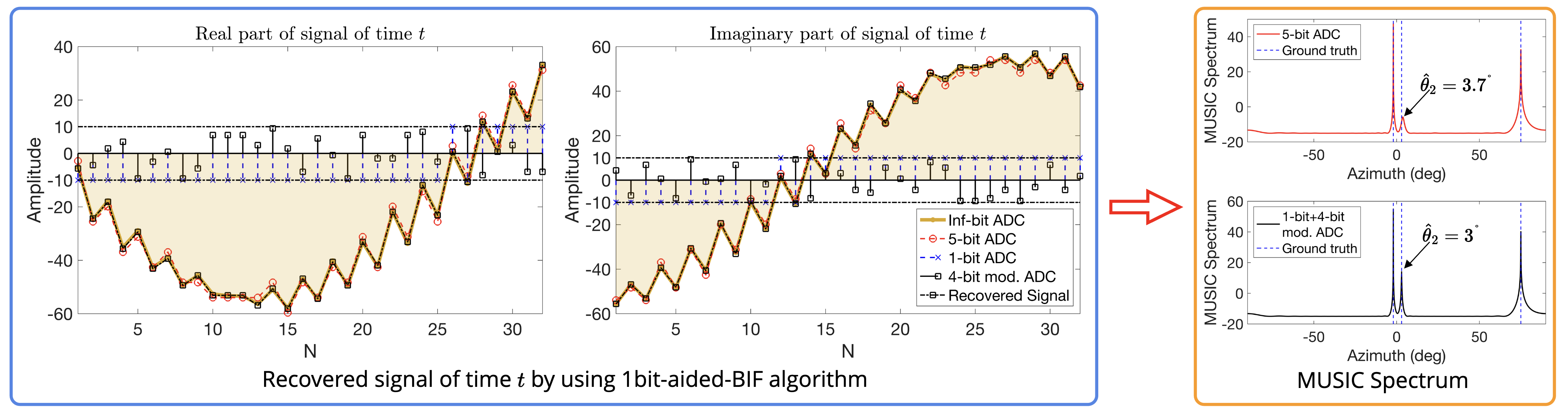}
		\caption{The performance of 1bit-aided-BIF algorithm in near-far problem with ${\rm SNR}_2 = -10$ dB and $T = 10^4$. }
		\label{simuDanci}
\end{figure*}
\begin{figure}[htb!]
		\centering
		\includegraphics[width=8.5cm]{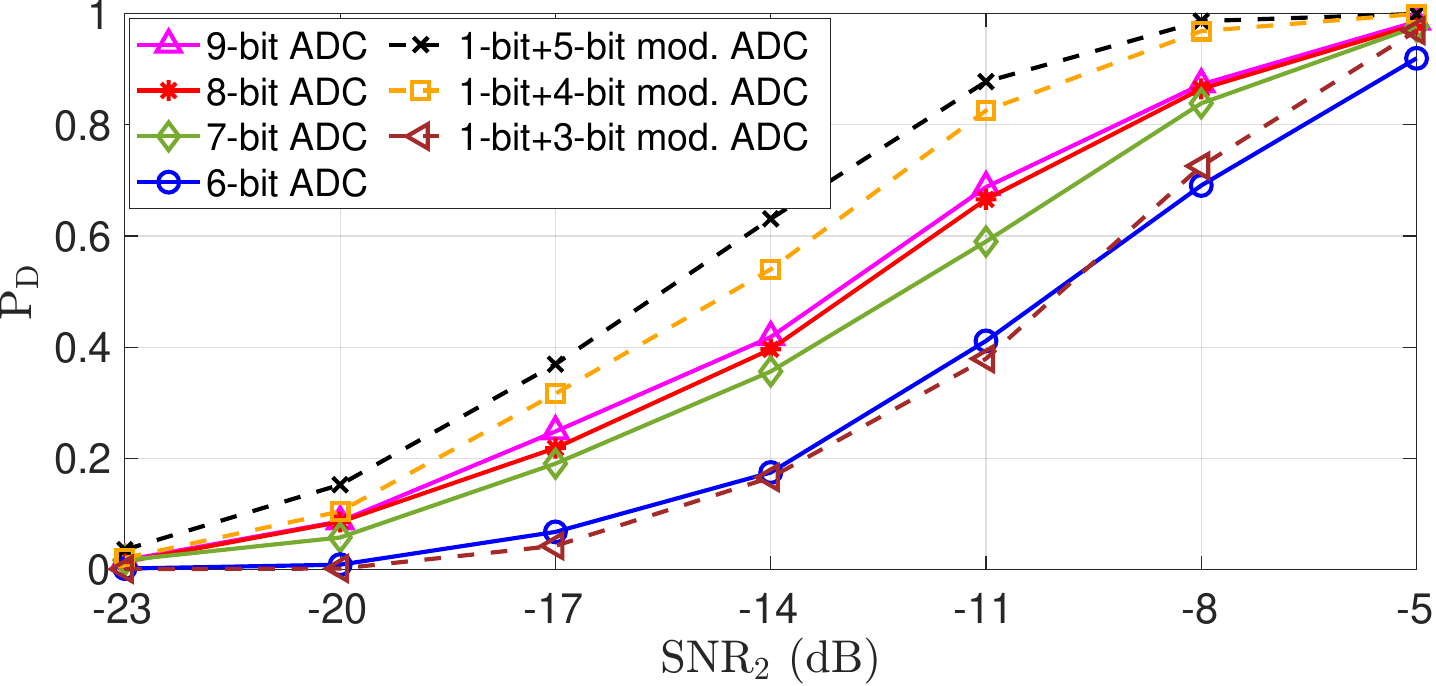}
		\caption{The detection probability versus ${\rm SNR}_2$.}
		\label{simuSNR}
\end{figure}

\begin{figure}[htb!]
		\centering
		\includegraphics[width=8.5cm]{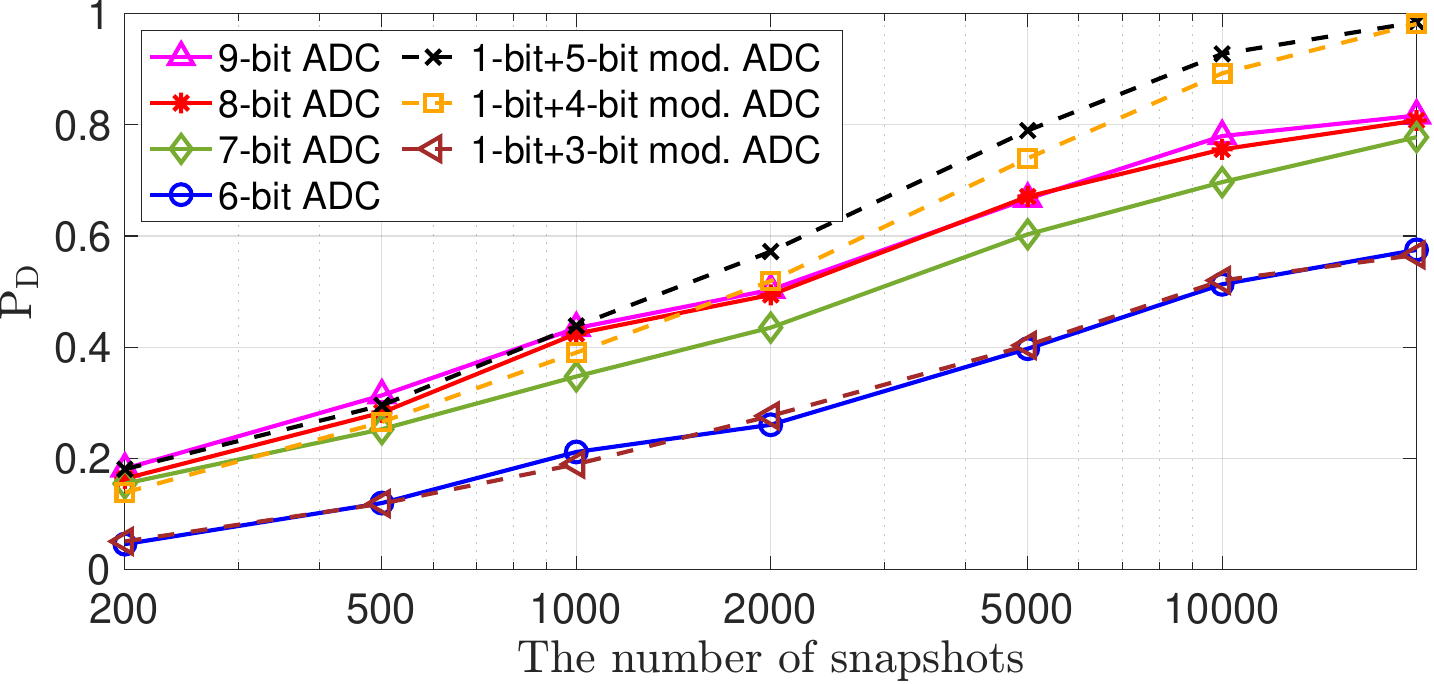}
		\caption{The detection probability versus the snapshot.}
		\label{simuL}
\end{figure}
\begin{algorithm}[htb!]
\caption{1bit-aided-BIF Algorithm}\label{1bitUSDOA_alg}
\begin{algorithmic}[1]
\STATE \textbf{Inputs:} Modulo samples ${\mathbf y}(t)$, $1$-bit samples $\mathbf{h}(t)$, $t=1,\cdots, T$; number of signals $K$; maximum number of iterations ${\rm Iter}_{\rm max}$.
\STATE Estimate the normalized covariance matrix ${\widetilde{\mathbf{C}}}_r$ according to the arcsine law and (\ref{norconmat}).
\STATE Initialize the estimate of the integer matrix $\mathbf{A}$ by solving (\ref{IFoptinit}) using the LLL algorithm.
\STATE Initialize the estimate of covariance matrix $\mathbf{C}_r$ according to (\ref{estcon}).
\WHILE{$\widehat{\mathbf{C}}_r$ does not converge or ${\rm Iter}_{\rm max}$ is not reached}
\STATE Calculate $\widehat{\mathbf{A}}$ by solving (\ref{IFopt}) using the LLL algorithm, where $\mathbf{C}_r$ is substituted with $\widehat{\mathbf{C}}_r$.
\STATE Update the estimate of the covariance matrix $\widehat{\mathbf{C}}_r$ (\ref{estcon}).
\ENDWHILE
\STATE Estimate the covariance matrix $\mathbf{C}$ (\ref{estcon2}).
\STATE Estimate the DOAs ${\boldsymbol{\theta}}$ by performing the root MUSIC on the estimated covariance matrix $\widehat{\mathbf{C}}$.
\STATE \textbf{Outputs:} $\{\widehat{\boldsymbol{\theta}}\}$.
\end{algorithmic}
\end{algorithm}
\section{Simulation}
\label{sec:simu}

In this section, numerical experiments are conducted using a ULA to verify the performance of the proposed 1bit-aided-BIF algorithm in the near-far problem. The number of sources is $K=3$, which are located at $\boldsymbol{\theta} = [-2^{\circ};3^{\circ};75^{\circ}]$. The signal-to-noise ratio (SNR) of the $k$-th source is defined as ${\rm SNR}_k \triangleq 20\log\sigma_k/\sigma$. We set ${\rm SNR}_1 = 30$ dB and ${\rm SNR}_3 = 15$ dB. The second signal $\theta_2$ will be set as the weak signal with lowest SNR in the following experiments. $\boldsymbol{\theta}$ is detected provided that $\min\{|\theta_k - \widehat{\theta}_i|\}_{i=1}^K \leq 0.1^{\circ}$ for all $k$, and the detection probability ${\rm P}_{\rm D}$ is used as a criterion for performance evaluation. For comparison with the benchmark (conventional ADC) and to highlight the benefits of modulo ADC in quantization, we assume that the dynamic range of conventional ADC matches the signal $\mathbf{g}(t)$, and the threshold of conventional ADC for the I and Q channels is set to four times the standard deviation, i.e., $4\sqrt{(\sum_{k=1}^K\sigma_k^2)/{2}}$. The proposed algorithm also addresses the case of sparse linear arrays. All statistical results presented here are averaged over $1000$ Monte Carlo trials.

Fig. \ref{simuDanci} shows the MUSIC spectrum of the 1bit-aided-BIF algorithm with a $4$-bit modulo ADC, and the samples and the recovered signal for $t=15$ are plotted. The signal of $t=15$ is perfectly recovered and has a smaller NMSE ($-41.3$ dB) compared to the samples obtained by 5-bit ADC ($-26.4$ dB) due to low quantization noise. Compared to conventional ADC, the MUSIC spectrum of the one-bit-aided modulo ADC framework results in improved DOA estimation performance, i.e., the peak near the weak target location ($3^\circ$) is sharper, and the corresponding peak localization is closer to the true DOA.

Fig. \ref{simuSNR} shows the performance of the 1bit-aided-BIF algorithm with different modulo ADC bit depths against the SNR of the weak signal. We set $T = 10^4$. In this simulation, the US-ASP method in \cite{fernandez2022computational} with $4$-bit, $5$-bit, $6$-bit modulo ADCs are performed for comparison and we set $d = \frac{\nu}{10}$ so that the spatial oversampling factor is $5$.\footnote{The parameters are defined in the same manner as in \cite{fernandez2022computational}.} Although a recovery guarantee is achieved for US-ASP in the absence of noise, US-ASP is sensitive to noise. In our noisy scenario, the detection probability of US-ASP is nearly zero; therefore, its results are not presented in Fig. \ref{simuSNR}. As SNR$_2$ increases, all cases show improved detection probabilities. Modulo ADC with $B=3$ (equivalent to $4$ bits, including the $1$-bit ADC) exhibits performance on par with a $6$-bit ADC. Furthermore, modulo ADCs with $B=4$ and $B=5$ (with totally $5$ and $6$ bits) outperform the $9$-bit ADC. 

Fig. \ref{simuL} depicts the 1bit-aided-BIF algorithm's performance across various modulo ADC bit depths versus the number of snapshots, with ${\rm SNR}_2 = -10$ dB. As $T$ increases, all cases exhibit enhanced detection probabilities. Specifically, modulo ADC with $B=3$ matches the performance of a $6$-bit ADC. Additionally, modulo ADCs with $B=4$ and $B=5$ (a total of $5$ and $6$ bits) surpass the $9$-bit ADC when $T>10^3$. As $T$ increases, the detection probabilities of modulo ADCs with $B=4$ and $B=5$ approach $1$.
\section{Conclusion}
\label{sec:con}
This paper proposes the one-bit-aided modulo ADC sampling framework to tackle the clipping and near-far problem in DOA estimation by combining one-bit sampling and modulo sampling. In addition, the 1bit-aided-BIF algorithm is proposed to perform the DOA estimation. Numerical experiments demonstrate the excellent performance of the proposed sampling architecture and algorithm compared to conventional high-precision ADC.

\end{document}